\documentclass[letter]{aa} 

\usepackage{graphicx}
\usepackage{natbib}
\bibpunct{(}{)}{;}{a}{}{,}

\usepackage{txfonts}
\usepackage{caption}
\usepackage{subcaption}
\usepackage{graphicx}
\usepackage{multirow}

\newcommand{\ci}{\mbox{[C{\small I}]}}

\begin{document} 

   \title{Molecular gas properties of Q1700-MD94: a massive, main-sequence galaxy at $z\approx2$}

   \author{K. Henr\'iquez-Brocal,
          \inst{1}
          R. Herrera-Camus \inst{1,2},
          L. Tacconi \inst{2},
          R. Genzel \inst{2},
          A. Bolatto \inst{3},
          S. Bovino \inst{1},
          R. Demarco \inst{1},
          N. Förster Schreiber \inst{2},
          M. Lee \inst{1},
          D. Lutz \inst{1},
          \and
          M. Rubio \inst{4}
          }
         \authorrunning{Brocal et al.}

   \institute{Departamento de Astronom\'ia, Facultad de Ciencias F\'isicas y Matem\'aticas, Universidad de Concepci\'on, Av. Esteban Iturra s/n Barrio Universitario, Casilla 160, Concepci\'on, Chile \label{1}
    \and
    Max-Planck-Institut f\"ur extraterrestische Physik (MPE), Giessenbachstr., D-85748 Garching, Germany
	\label{2}
	\and
	Department of Astronomy, University of Maryland, 
	College Park, MD 20742, USA 
	\label{3}
	\and
	 Departamento de Astronomía, Universidad de Chile, Casilla 36-D Santiago, Chile
	\label{4}
             }


 
  \abstract
   {
   We use a combination of new NOrthern Extended Millimeter Array (NOEMA) observations of the pair of \ci\ transitions, the CO(7-6) line, and the dust continuum, in addition to ancillary CO(1-0) and CO(3-2) data, to study the molecular gas properties of Q1700-MD94, a massive, main-sequence galaxy at $z\approx2$. We find that for a reasonable set of assumptions for a typical massive star-forming galaxy, the CO(1-0), the \ci(1-0) and the dust continuum yield molecular gas masses that are consistent within a factor of $\sim2$. The global excitation properties of the molecular gas as traced by the \ci\ and CO transitions are similar to those observed in other massive, star-forming galaxies at $z\sim2$. Our large velocity gradient (LVG) modeling using RADEX of the CO and \ci\ spectral line energy distributions (SLEDs) suggests the presence of relatively warm ($T_{\rm kin}=41$~K), dense ($n_{\rm H_2}=8\times10^{3}~{\rm cm}^{-3}$) molecular gas, comparable to the high-excitation molecular gas component observed in main-sequence, star-forming galaxies at $z\sim1$.
   The galaxy size in the CO(1-0) and CO(7-6) line emission are comparable, which suggests that the highly-excited molecular gas is distributed throughout the disk powered by intense star formation activity. To confirm this scenario will require spatially resolved observations of the CO and \ci\ lines which can now be obtained with NOEMA upgraded capabilities.
   }

   
\maketitle

\section{Introduction}

Understanding galaxy evolution requires the study of the complex interplay between multiple physical processes. One of the most important ones is star formation, which is connected to the reservoir of molecular gas (H$_2$) available, and the physical conditions that set the efficiency at which the gas is converted into stars \cite[e.g.,][]{kennicutt12}. Despite its importance, measuring the amount of molecular gas in galaxies across cosmic time remains a challenge \citep[e.g.,][]{tacconi2020}. The most common tracer of H$_2$ gas is the first rotational transition of CO ($J=1-0$), which becomes progressively harder to observe as the redshift of the source increases \citep[e.g.,][]{carilli13}. In addition, to convert the CO(1-0) luminosity into a molecular gas mass it is necessary to assume a conversion factor ($\alpha_{\rm CO}$), which is strongly dependent on the metallicity, as in low extinction regions the interstellar radiation field can strongly photodissociate CO molecules \citep[e.g.,][]{bolatto2013}.

The pair of neutral carbon (C) fine-structures transitions, \ci(1-0) at 492.161 GHz and \ci(2-1) at 809.344 GHz, has emerged as an alternative to the use of the CO lines as a tracer of the molecular gas, in particular for high redshift galaxies where the \ci\ lines shift into the millimeter bands. In contrast to what is expected from the classical onion-like structure of photodissociation regions \citep[PDRs; e.g.,][]{hollenbach1999}, more recent simulations suggest that neutral carbon emission appears to be widespread in H$_2$ clouds \citep[e.g.,][]{offner2014,glover2016,bisbas2017}. Observations that support this scenario include large scale maps of the CO isotopologues and \ci\ lines of star-forming regions in the Small and Large Magellanic Clouds \citep{requena2016,okada2019}, and the $\rho$~Ophiuchi \citep{kulesa2005} and Orion giant molecular clouds. For the latter, the \ci(1-0) emission is measured to be optically thin ($\tau_{\rm [CI]}\sim 0.1 - 0.3$), and coincident with the emission of the $^{12}$CO line and $^{13}$CO(1-0) lines \citep[][]{ikeda2002,shimariji2013}. In nearby star-forming galaxies and starbursts, kiloparsec scale observations suggest that \ci\ may be as good a tracer of the cold H$_2$ gas as CO, but not superior \citep[e.g.,][]{israel2015,crocker2019}. 

With the advent of ALMA and the boosted capabilities of IRAM/NOEMA, the number of galaxies at high redshift observed in the \ci\ lines is rapidly increasing \citep[e.g.,][]{bothwell2017,Lee2021}, and observations of the \ci\ lines are no longer restricted to  bright quasars or sub-millimeter galaxies \citep[e.g.,][]{weiss2003,weiss2005,walter2011}. In the first systematic study of the \ci\ line emission of main-sequence galaxies and starbursts at $z\gtrsim1$, \cite{valentino2018,valentino2020_CI} find that the \ci(1-0) transition is a good tracer of the molecular gas mass, similar to low-$J$ CO lines and dust, and that the gas temperature traced by the \ci\ lines do not strongly vary across galaxy types and redshift. 

In combination with multiple CO transitions, the pair of \ci\ lines can be used to constrain the physical properties of the molecular gas. \cite{valentino2020}, based on Large Velocity Gradient (LVG) modeling of the CO+\ci\ spectral line energy distributions (SLEDs), find that to reproduce the observed SLEDs of $z\sim1$ main-sequence galaxies at least two molecular gas components are necessary: one that is relatively diffuse ($n_{\rm H_2}\approx10^2$~cm$^{-3}$, $T_{\rm kin}\approx45$~K) and one that is dense ($n_{\rm H_2}\approx10^4$~cm$^{-3}$, $T_{\rm kin}\approx45$~K). In a similar analysis of BX610, a massive, main-sequence galaxy at $z\approx2$, \cite{brisbin2019} also find that two molecular gas components are an option, with the high density one heated either by intense star formation or low-velocity C shocks. There is certainly much more work to be done to constrain the molecular gas properties of galaxies across cosmic time, and the combination of CO+\ci\ SLEDs is one promising tool to achieve this goal. 

In this Letter, we add to the limited number of studies of the molecular gas properties of main-sequence galaxies at high-redshift based on observations of the pair of \ci\ lines, CO(1-0) and a couple of mid-$J$ CO lines. The subject of study is Q1700-MD94 (hereafter MD94), a massive ($M_{\star}=1.5\times10^{11}~M_{\odot}$), star-forming (${\rm SFR}=271~M_{\odot}~{\rm yr}^{-1}$) system at $z=2.333$ \citep{tacconi2013} located at the massive end of the main-sequence of galaxies at $z\sim2$ \citep[e.g.,][]{whitaker12}. MD94 is located at a declination of +64d 11m 24.2s, and has been previously observed in CO(1-0) \citep{bolatto2015} and CO(3-2) line emission as part of the PHIBSS survey \citep{tacconi2013}.


\section{Observations}

We observed MD94 with NOEMA located in the French Alps. On August 31, 2017, we used the 2~mm band for a total of 7.5 hours to target the \ci(1-0) line located at an observed frequency of $\nu_{\rm obs}=147.8$~GHz. Then, starting on September 1, 2017, we used the 1 mm band for a total of 14.3 hours to observe the \ci(2-1) and CO(7-6) lines centered at an observed frequency of $\nu_{\rm obs}=243.0$~GHz. All observations were carried out in NOEMA D array configuration, which is the most compact one.

The data were calibrated in CLIC with help from the staff in Grenoble. After calibration, we used the software MAPPING2\footnote{CLIC and MAPPING2 are part of the GILDAS package \citep{guilloteau2000}. http://www.iram.fr/IRAMFR/GILDAS} to clean and image the data using natural weighting. For the \ci(1-0) cube the resulting synthesized beam was $3.0'' \times 2.2''$ ($\sim24.9\times18.2$~kpc) and the noise level was $0.25~\rm mJy~beam^{-1}$ in 100~km~s$^{-1}$. For the \ci(2-1) and CO(7-6) cube the resulting beam was $1.7'' \times 1.3''$ ($\sim14.1\times10.7$~kpc) and the noise level was $0.31~\rm mJy~beam^{-1}$ in 75~km~s$^{-1}$. In the latter cube we detected continuum emission, which was subtracted from the spectra before any line analysis was done. 

\begin{figure*}
\includegraphics[width=\hsize]{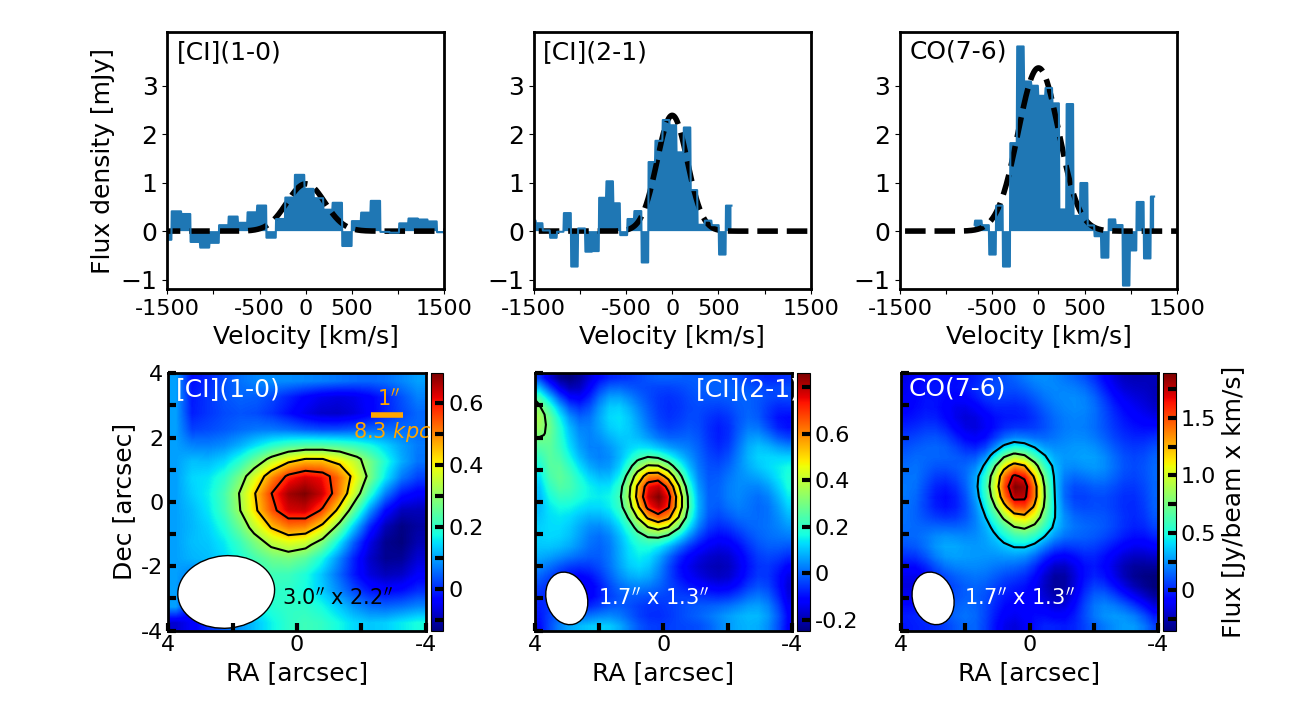}
\caption{{\it (Top)} From left to right, \ci(1-0), \ci(2-1), and CO(7-6) spectra of MD94. The black-dashed lines show the single Gaussian fit to the spectra. The FWHM and integrated fluxes are listed in Table~1. {\it (Bottom)} From left to right, integrated intensity maps of the \ci(1-0), \ci(2-1), and CO(7-6) emission of MD94. Contours start at $2\sigma$ and increase in steps of $1\sigma$. The NOEMA synthesized beam is shown in the lower left corner.}\label{fig:moment}
\end{figure*}

\section{Results}

\subsection{Detection of the \ci(1-0), \ci(2-1) and CO(7-6) transitions and the dust continuum} 

The spectra of the \ci(1-0), \ci(2-1) and CO(7-6) lines observed in MD94 are shown in the top panel of Fig.~\ref{fig:moment}. The continuum emission detected under the \ci(2-1) and CO(7-6) lines was determined and subtracted based on the emission from the frequency channels below and above the CO(7-6) and \ci(2-1) lines, respectively. The continuum flux measured at (rest-frame) 240~GHz is $2.5\pm0.8$~mJy. The \ci(1-0), \ci(2-1) and CO(7-6) lines are detected with signal-to-noise ratios of $S/N\approx4,~ 6~{\rm and} ~10$, respectively. We fit a single Gaussian profile to each line. The measured linewidths (FWHM) and integrated fluxes are listed in Table~1. The redshift of the source derived from the central frequency of the Gaussian fit is similar to the value of $z=2.33$ measured from observations of the H$\alpha$ and CO(3-2) lines by \cite{tacconi2013}.

The bottom panels of Fig.~\ref{fig:moment} show the integrated intensity maps. We (barely) spatially resolve MD94 in the \ci(2-1) and CO(7-6) line emission. Similar to the spatial distribution observed in the CO(1-0) line emission by \cite{bolatto2015}, the source is elongated in the N-S direction. The deconvolved sizes are $1.4''\times0.7''$ and $1.2''\times0.61''$ for the \ci(2-1) and CO(7-6) lines, respectively. These sizes are comparable to the deconvolved source size of $1.3''\times1.0''$ measured from the CO(1-0) line emission \citep{bolatto2015}.

We calculated the luminosities of the \ci\ and CO lines following \cite{solomon2005}. For the luminosity distance and other physical parameters we use a standard cosmology \citep[$H_0=67.4$~km~s$^{-1}$~Mpc$^{-1}$, $\Omega_{\rm M}=0.315$, $\Omega_{\Lambda}=0.685$][]{planck2020}.

\begin{table}[ht]
\caption{Q1700-MD94 derived values} 
\centering 
\begin{tabular}{c c c c} 
\hline\hline 
Line & FWHM & $S\Delta v$ & Reference\\ [0.5ex] 
& [km s$^{-1}]$ & [Jy km s$^{-1}$] &  \\
\hline
\ci(1-0) & $482\pm124 $ & $0.49 \pm 0.13$ & This work\\
\ci(2-1) & $383\pm75$ & $0.92 \pm 0.15$ & This work\\
CO(1-0) & $296\pm66 $ & $0.14 \pm 0.01$ & \cite{bolatto2015} \\ 
CO(3-2) & $294\pm49$ & $1.52 \pm 0.17$ & \cite{bolatto2015} \\ 
CO(7-6) & $518\pm 61 $ & $1.67 \pm 0.17$ & This work \\ 
\hline
Continuum & & Flux density & \\ [0.5ex] 
&  & [mJy] &  \\
\hline
$S_{240~\rm GHz}$ &  & $2.5\pm0.8$ & This work \\ 
\hline 
\end{tabular}
\label{table:data_table1} 
\end{table}

\subsection{The molecular gas mass of MD94}

The molecular gas mass of MD94 can be measured from available observations of the CO(1-0) line \citep{bolatto2015}, the dust continuum, and the pair of \ci\ lines (this work). In all cases it is necessary to make assumptions about the abundance of the tracer to convert the luminosity into a molecular gas mass. 

According to the mass-metallicity relation and parameterization in \cite{genzel2015},MD94 has a metallicity of $12 + \rm log(O/H) = 8.64$, which is roughly the solar value \citep{asplund2004}. Following \cite{tacconi18}, for the metallicity of MD94 the CO(1-0) metallicity dependent conversion factor is $\alpha_{\rm CO} = 4.5$ $M_{\odot}$ (K km $\rm s^{-1} \rm pc^2)^{-1}$ (this includes the 1.36 helium correction factor). Based on this, and the CO(1-0) luminosity in \cite{bolatto2015}, the molecular gas mass in MD94 is $M_{\rm mol}=1.7
\times10^{11}~M_{\odot}$. 

Alternatively, the interstellar medium (ISM) gas mass can be estimated from the single dust continuum measurement at (rest-frame) 240~GHz (see Table~\ref{table:data_table1}). Following Eq. (15) in \cite{scoville2016}, we measure a gas mass in MD94 of $M_{\rm ISM}\approx7\times10^{10}$~$M_{\odot}$ \footnote{This calculation assumes a dust-to-gas ratio appropriate for  massive, star-forming galaxies, and  includes a Rayleigh–Jeans correction calculated for a dust temperature of $T=35$~K. This temperature is consistent with the gas temperature derived from the ratio of the \ci\ lines in MD94, and also the typical dust temperature measured in massive, main-sequence galaxies at $z\approx2$ \citep[e.g.,][]{magnelli14,genzel2015,schreiber18}}. This total gas mass is within factor of $\sim2$ the CO-based molecular gas mass, which is expected given the multiple assumptions both methods require.

Finally, the atomic carbon mass based on the \ci(1-0) line can be calculated following, e.g., \cite{weiss2005}:

\begin{equation}
M_{\rm [CI]} = 5.706 \times 10^{-4} Q( T_{\rm ex})\frac{1}{3} {\rm e}^{\frac{23.6}{T_{\rm ex}}} L^{'}_{\rm [CI](1-0)}~M_{\odot},     
\end{equation}

\noindent where $T_{\rm ex}$ is the excitation temperature and $Q( T_{\rm ex})=1+3{\rm e}^{-T_{1}/T_{\rm ex}} + 5{\rm e}^{-T_{2}/T_{\rm ex}}$ is the partition function. Here $T_1=23.6$~K and $T_2=62.5$~K are the energies above ground state.

The advantage of having detected the pair of \ci\ lines in MD94, is that we can use their brightness ratio to directly measure the excitation temperature $T_{\rm ex}$ in the optically thin limit. In MD94 we measure a \ci\ line ratio of $R=L^{'}_{\rm [CI](2-1)}/L^{'}_{\rm [CI](1-0)}=0.7$. Following \cite{stutzki97}, the excitation temperature in MD94 is

\begin{equation}
T_{\rm ex} = 38.8\times\ln{\left(\frac{2.11}{\rm R}\right)}^{-1} \approx 35~{\rm K}.
\end{equation}

\noindent This value is close to the average \ci\ excitation temperature of $T_{\rm ex}\approx30$~K measured in the sample of dusty starbursts and QSOs in \cite{walter2011}, and about $\sim$10~K warmer than the average excitation temperature measured in main-sequence galaxies at $z\sim1$ \citep{valentino2020_CI}. In the absence of one of the two \ci\ lines, it is a common approach to assume the dust temperature as a proxy for the excitation temperature. In the case of MD94, the \ci-based $T_{\rm ex}$ is comparable to the typical dust temperature of $T_{\rm dust}\approx32$~K measured in massive ($M_{\star}\approx10^{11}-10^{11.5}~M_{\odot}$), main-sequence star-forming galaxies at $z\approx2$ \citep{magnelli14,genzel2015,schreiber18}.

From Eq.~(1), and the excitation temperature in Eq.~(2), the resulting atomic carbon mass in MD94 is $M_{\rm [CI]}=9.2\times10^{6}~M_{\odot}$. To convert this quantity into a molecular gas mass we need to assume an atomic carbon abundance relative to H$_2$, i.e., $X_{\rm CI}=M_{\rm CI}/6M_{\rm H_{2}}$. There is a wide range of $X_{\rm CI}$ values indirectly measured or assumed for different galaxy types and cosmic epochs. \cite{valentino2018}, using a homogenized method to compare the molecular gas masses derived from archival dust, low-$J$ CO, and \ci(1-0) line observations, find that $X_{\rm CI}$ varies from $\sim1-3\times10^{-5}$ in $z\sim1$ main-sequence galaxies, $\sim3-8\times10^{-5}$ in SMGs at $z\sim2$, and $\sim3-10\times10^{-5}$ in nearby galaxies (note that $M_{\rm H_2}$ values in Valentino et al. (2018) do not include the helium contribution). 
For MD94, if we compare the atomic carbon mass to the average molecular gas mass from the CO(1-0) line and dust continuum estimates, we find $X_{\rm CI,MD94}\approx1.3\times10^{-5}$. This value is within the range of abundances derived by \cite{valentino2018}, and comparable to the typically assumed abundance for massive, high redshift systems of $X_{\rm CI}\approx3\times10^{-5}$  \citep[e.g.,][]{weiss2005,Zadeh2013,popping17}. In summary, for reasonable assumptions on $\alpha_{\rm CO}$, dust-to-gas ratio, and $X_{\rm CI}$ for a massive, near solar metallicity galaxy like MD94, we find that the molecular gas mass measurements based on the CO(1-0), dust continuum, and \ci(1-0) transitions are comparable (e.g., within a factor of $\sim2$).

Finally, the molecular gas mass in MD94 implies this galaxy has a molecular gas fraction of $\mu=M_{\rm mol}/M_{\star}\approx1$ and a molecular gas depletion timescale of $t_{\rm dep}=M_{\rm mol}/{\rm SFR}\approx0.5$~Gyr. Both quantities are characteristic of massive, main-sequence galaxies at $z\sim2$ \citep{tacconi18}. 

\begin{figure*}
    \includegraphics[width=\hsize]{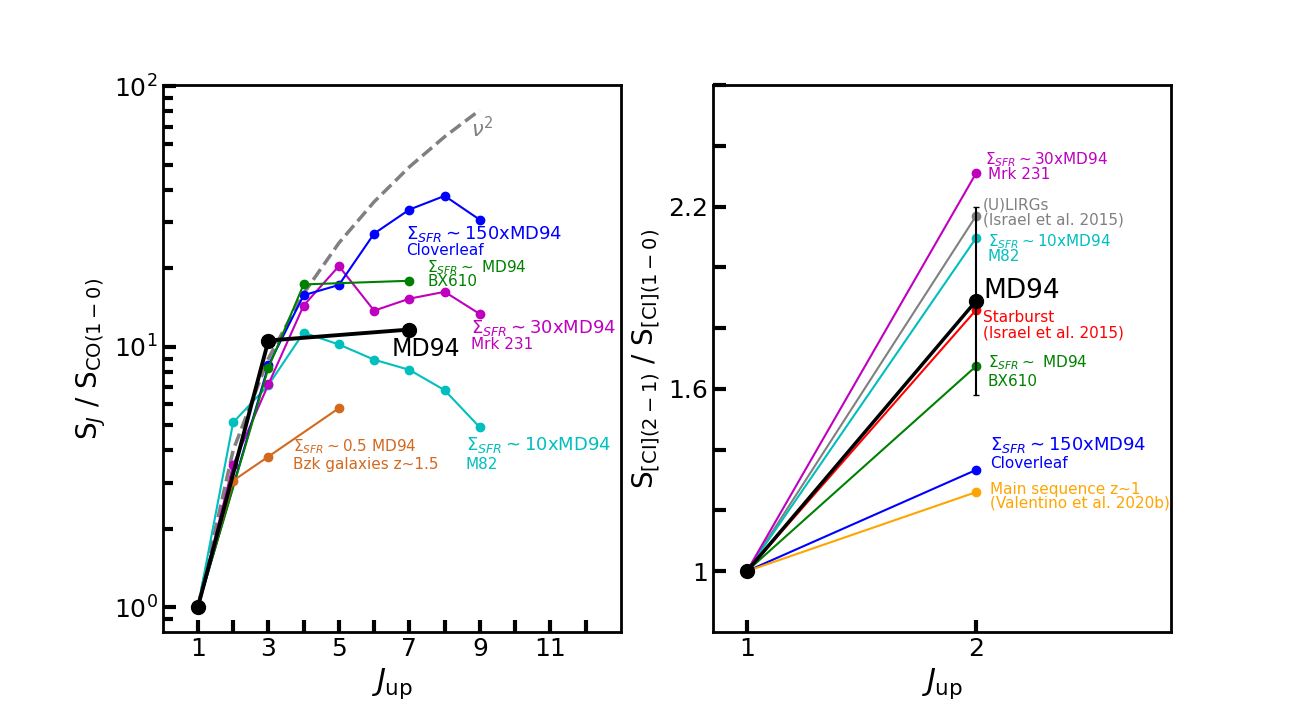} 
    \label{fig:sled}
    \caption{CO ({\it left}) and \ci\ ({\it right}) SLEDs of MD94 (black), and other nearby and high-redshift systems. We indicate (when available) the SFR surface density in terms of the value measured in MD94 ($\Sigma_{\rm SFR}\approx2~M_{\odot}$~yr$^{-1}$~kpc$^{-2}$). 
    The references for the ancillary data shown are the following: Mrk~231 \citep{vanderwerf10}, M82 \citep{kamenetzky2012}, and Cloverleaf \citep{barbainis1997}. In addition to these, we added BX610 \citep{brisbin2019,aravena2014}, average values for   (U)LIRGs and starbursts \citep{israel2015}, and the mean value for main-sequence galaxies at $z\sim1$ \citep{valentino2020_CI}, and BzK main-sequence galaxies at $z\sim1.5$ \citep{Daddi15}. The dashed dark gray line shows constant brightness temperature on the Rayleigh–Jeans scale ($S\propto \nu^2$).}\label{fig:sled}
\end{figure*}

\subsection{\ci\ and CO Spectral Line Energy Distributions}

The analysis of the SLED of the pair of \ci\ and CO lines can provide valuable insights into the physical conditions of the molecular gas. Fig.~\ref{fig:sled} shows the CO and \ci\ SLEDs of MD94 compared to other nearby and high-redshift galaxies. 
Given the strong link observed between the shape of the SLED and the SFR surface density $\Sigma_{\rm SFR}$ \citep[e.g.,][]{narayanan2014}, we include galaxies that have $\Sigma_{\rm SFR}$ values that range from $\sim1\times$ (e.g., BX610, mean BzK $z\sim1.5$ galaxies) to $\sim150\times$ (e.g., Cloverleaf quasar) that of MD94 \citep[$\Sigma_{\rm SFR}\approx2~M_{\odot}$~yr$^{-1}$~kpc$^{-2}$;][]{tacconi2013}.

The integrated flux ($S_{\rm CO}\Delta v$) ratio between the $J=3-2$ and $J=1-0$ in MD94 is $\sim10$ \citep{bolatto2015}, 
which is consistent with its high levels of star formation activity, similar to the examples of M82 (in the figure) and NGC253 \citep{bradford2003}. It is also consistent with the high integrated line ratios observed in AGN galaxies, such as Mrk~231 and the Cloverleaf quasar (in the figure). It is important to note that, from the observed broad H$\alpha$ line profile \citep{erb2006}, MD94 could host an AGN possibly accompanied by an outflow. 


The inclusion of the CO(7-6) line is key to further differentiate the gas excitation conditions in MD94 from other systems, and also to search for an additional denser and/or warmer component of molecular gas present. In that sense, the (partially complete) SLED of MD94 from $J=1$ to $J=7$ better resembles the shape of the SLED of nearby AGN galaxy Mrk~231 ($\Sigma_{\rm SFR}\sim30\times\Sigma_{\rm SFR,MD94}$), than the nearby starburst M82 ($\Sigma_{\rm SFR}\sim10\times\Sigma_{\rm SFR,MD94}$) that peaks at $J=4$ and then declines. It is also similar to the SLED of BX610, another massive, main-sequence galaxy at $z\approx2$ with a comparable value of $\Sigma_{\rm SFR}$ and showing no or weak signs of AGN activity \citep{aravena2014,bolatto2015,brisbin2019}. For the latter, LVG models suggest that at least a second, higher-density PDR component or a low-velocity C shock heated component is needed to better reproduce the CO+\ci\ SLED \citep{brisbin2019}. 


The right panel of Fig.~\ref{fig:sled} shows the \ci\ SLED of MD94. We also include most of the systems in the left panel plus the average \ci\ SLED observed in main-sequence galaxies at $z\sim1$ \citep{valentino2018} and nearby starbursts and (U)LIRGs \citep{israel2015}. MD94 has a \ci(2-1)/\ci(1-0) integrated flux ratio of $\approx1.9$, similar to the typical value found in nearby starbursts \citep{israel2015}, and slightly higher than BX610 \citep{brisbin2019}. The fact that MD94 shows a \ci\ SLED that is more excited than, for instance, the Cloverleaf QSO ($\Sigma_{\rm SFR}\sim150\times\Sigma_{\rm SFR,MD94}$), reveals the complexity of the molecular gas in these systems in terms of excitation properties and structure. It also suggests that the molecular gas component dominating the mid- and high-$J$ CO emission is different from that powering the \ci\ line emission. We further investigate the shape of the CO+\ci\ SLEDs in the next section. 

\begin{figure*}[t]
    \includegraphics[width=\hsize]{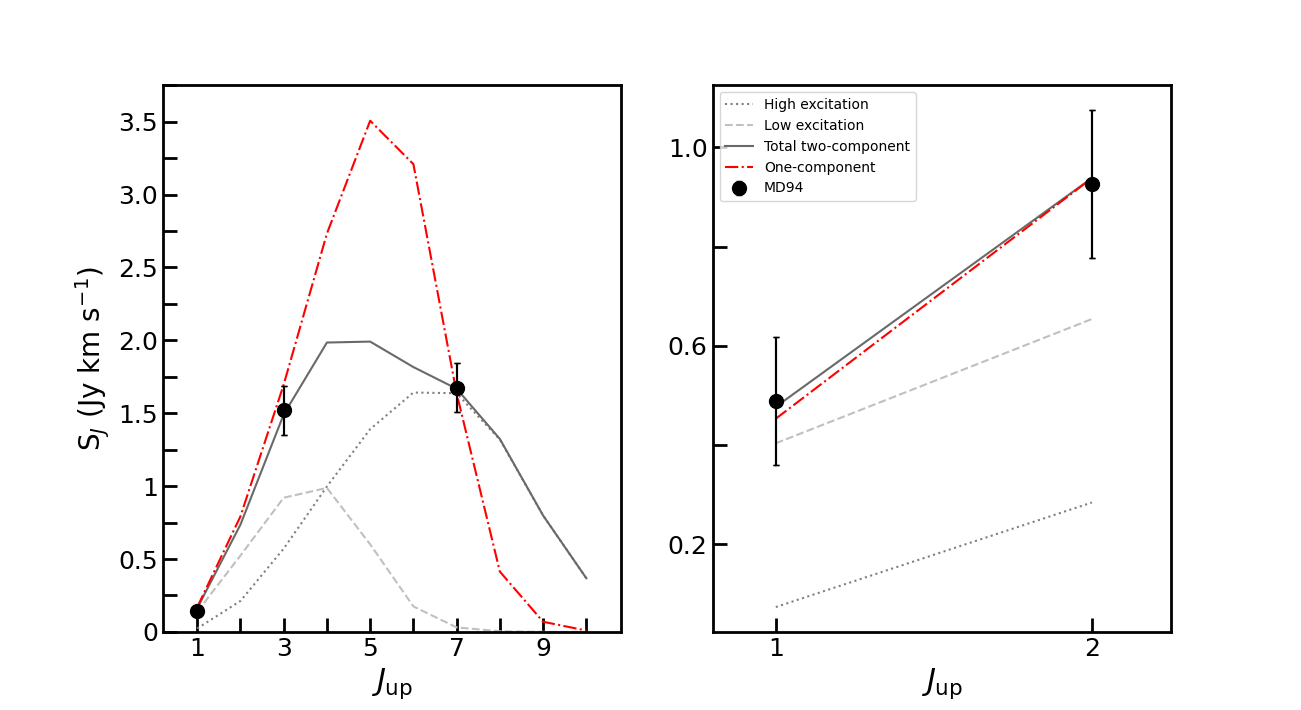} 
    \label{fig:twocomponent}
    \caption{LVG RADEX modeling of the CO (left) and \ci\ (right) SLEDs. The red line shows the best RADEX single-component model ($T=41$~K, $n_{\rm H_2}=8.5\times10^{3}~{\rm cm}^{-3}$). The gray line shows the best two-component model, which is the combination of a low- ($T=34$~K, $n_{\rm H_2}=3.1\times10^{3}~{\rm cm}^{-3}$; green) and a high-excitation ($T=137$~K, $n_{\rm H_2}=1.6\times10^{4}~{\rm cm}^{-3}$; red) components.
    }\label{fig:twocomponent}
\end{figure*}

\begin{figure}[t]
\includegraphics[width=\hsize]{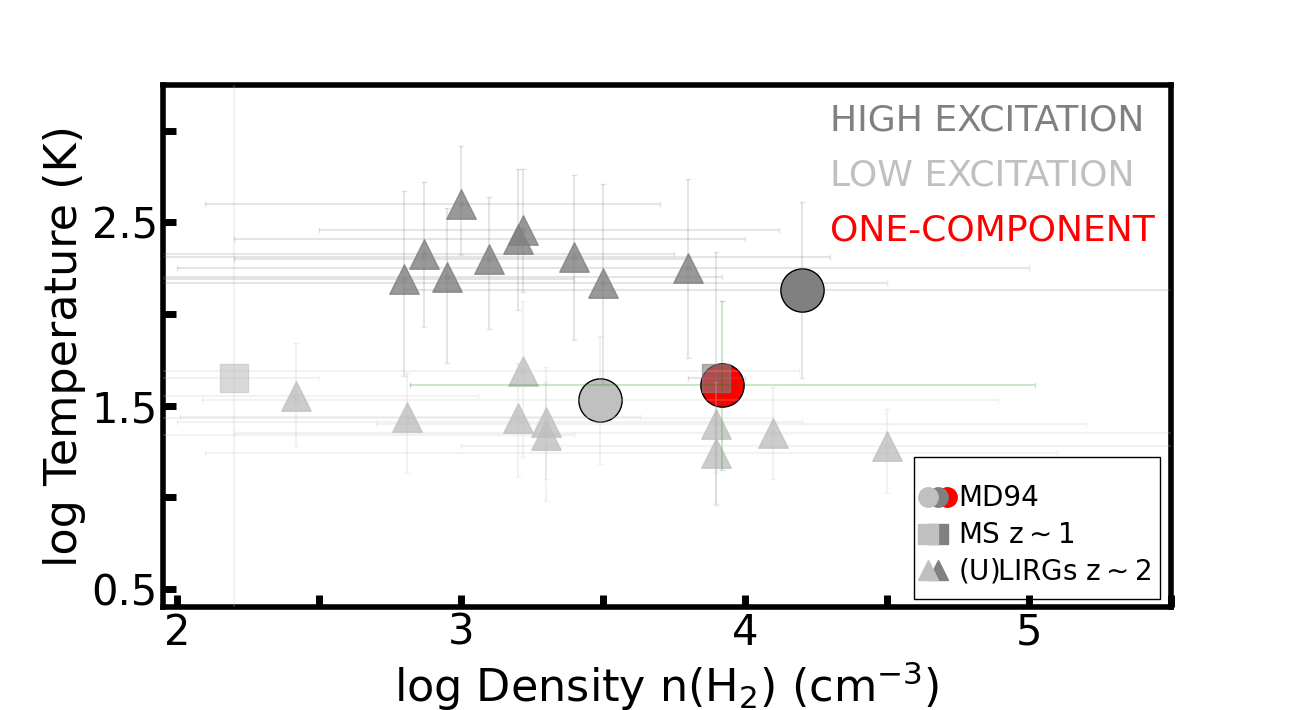}
  \caption{Kinetic temperature and molecular gas density from the best RADEX LVG one-component (red circle) and two-component model (dark gray and light gray circles) fitted to the CO and \ci\ SLEDs of MD94. We also show the derived values for lensed ULIRGs at $z\sim2-3$ \citep{yang2017,canameras2018} (triangles), and the average value for main-sequence galaxies at $z\sim1$ \citep{valentino2020} (squares).}\label{fig:tyn}
\end{figure}

\subsection{Large Velocity Gradient modeling}

To look deeper into the physical conditions of the gas that produce the CO and \ci\ line emission in MD94, we use a set of LVG models generated by the one-dimensional, non-LTE radiative transfer code RADEX \citep{vandertak2007}. RADEX models mainly depend on three input parameters: gas kinetic temperature ($T_{\rm K}$), molecular hydrogen density ($n_{\rm H2}$), and the CO (or C) column density per unit velocity gradient ($N({\rm CO})/dV$ or $N({\rm C})/dV$). To model the CO+\ci\ SLED of MD94 we generated a grid of models varying the H$_2$ density from $n_{\rm H_2}=10^2$ to $10^5$~cm$^{-3}$, and the kinetic temperature from $T_{\rm kin}=10$ to $200$~K. Following the LVG modeling of \cite{israel2015}, and to avoid overfitting the limited CO SLED of MD94, we fixed the CO gradient to $N({\rm CO})/dV=10^{17}$~cm$^{-2}~({\rm km~s}^{-1})^{-1}$, but note that varying $N({\rm CO})/dV$ in the range $10^{17}-10^{18}~({\rm km~s}^{-1})^{-1}$ only changes the results by less than 20\%. We also assumed an abundance ratio of C:CO of 1:4, consistent with abundance ratios found in nearby star-forming galaxies \citep{crocker2019} and the $z\sim2$ main-sequence galaxy BX610 \citep{brisbin2019}. The background temperature was set to $2.73\times(1+2.33)=9$~K.

We simultaneously compare the CO and \ci\ SLEDs of MD94 with the set of RADEX models, calculating the goodness of the fit based on the reduced-$\chi^2$ value. We find that the model that best reproduces the data has (reduced-$\chi^2\approx3$) has a gas density of $n_{\rm H_2}=8.5\times10^{3}~{\rm cm}^{-3}$ and a temperature of $T_{\rm kin}=41$~K. Fig.~\ref{fig:twocomponent} shows this best-fit single component model (blue dotted-dashed line) overplotted on the CO+\ci\ SLED of MD94. The model peaks between $J=4-6$, so future observations of the CO(4-3) or CO(5-4) transitions with NOEMA/Band 2, or CO(6-5) with NOEMA/Band 3, can be particularly helpful to better constrain the CO SLED model.

Among nearby galaxies and high redshift systems, it is not uncommon that a two-component model, consisting of the combination of low-excitation (or cold) and high-excitation (or warm) gas, is needed to fully reproduce the observed SLEDs \citep[e.g.,][]{mashian2015,Daddi15,canameras2018,valentino2020}. In the case of MD94, we are limited by the number of line transitions observed (only 5), so a proper fit based on a two-component model is out of reach. However, if for instructive purposes we fit the CO+\ci\ SLED with a two-component model, we find that the best model consists of a low-excitation component with $T_{\rm kin}=34$~K, $n_{\rm H_2}=3.1\times10^{3}~{\rm cm}^{-3}$ and a high-excitation component with $T_{\rm kin}=137$~K, $n_{\rm H_2}=1.6\times10^{4}~{\rm cm}^{-3}$. This two-component model is shown in Fig.~\ref{fig:twocomponent} as a combination of solid, dotted, and dashed gray lines.

\subsection{How does the physical properties of the molecular gas of MD94 compare to other galaxies at $z\sim1-2$}

There are two other main-sequence galaxies at $z\approx2$ with similar stellar mass as MD94 ($\sim10^{11}~M_{\odot}$) that have been studied based on the combination of the \ci\ and CO lines and the dust continuum. These galaxies are BX610 \citep{tacconi2013,bolatto2015,brisbin2019} and GS30274 \citep{popping17,talia18}. While the former share the average molecular gas properties of main-sequence galaxies at $z\sim2$ \citep{tacconi18}, the latter shows a very short depletion time ($t_{\rm dep}\approx100$~Myr), compact structure ($R_{\rm e}\approx2$~kpc), a powerful AGN-driven outflow \cite[e.g.,][]{nfs14,genzel14,davies20}, and it is considered to be in the process of quenching \citep{popping17}. 

Massive, main-sequence galaxies at $z\sim2$ are expected to experience star formation quenching as the fraction of quiescent galaxies with stellar masses similar to MD94 increases from $\sim20\%$ at $z\sim2-3$ to $\sim70\%$ at $z\sim1$ \citep[e.g.][]{muzzin2013}. From the standpoint of the global molecular gas fraction and depletion timescale, MD94 is closer to BX610 than GS30274, and does not show evident signatures for quenching of its star formation activity. This could be, of course, a result of the poor spatial resolution of our observations that washes out any evidence of quenching mechanisms operating in the central region around the AGN. The only similarity we find between MD94 and the quenching galaxy GS30274 is that the LVG modeling of MD94 suggests the presence of 
dense ($\sim8\times10^3~{\rm cm^{-3}}$) gas similar to that observed in GS30274. As Fig.~\ref{fig:tyn} shows, the gas temperature and density in MD94 is comparable to the high-excitation gas component in main-sequence galaxies at $z\sim1$ \citep{valentino2020}, and as dense as the molecular gas detected in highly-lensed starbursts at $z\sim2$ \citep{yang2017,canameras2018}. For MD94, this could be the result of highly-excited gas in the central region as a result, for example, of a compaction event. However, the fact that the deconvolved sizes of MD94 in the CO(1-0) and CO(7-6) line emission are comparable, suggests that the dense gas excited by intense star formation activity is well mixed with cold and diffuse molecular gas throughout the disk. Only high-spatial resolution observations of the molecular gas in multiple transitions, which can be achieved for example by NOEMA upgraded capabilities (e.g., baseline extension), could provide an answer in the near future. 


\section{Conclusions}

We present new NOEMA observations of the \ci(1-0), \ci(2-1) and CO(7-6) transitions of the massive, main-sequence galaxy Q1700-MD94 at $z\approx2.3$. Combined with ancillary observations of the CO(1-0) \citep{bolatto2015} and the CO(3-2) \citep{tacconi2013} lines, we have characterized the molecular gas properties of MD94. We highlight the following points:

\begin{itemize}

\item We find that for a reasonable set of assumptions for a massive galaxy such as MD94, the molecular gas mass estimates from the dust continuum, the CO(1-0), and the \ci(1-0) line are consistent within a factor of $\sim2$. At least for this system, our analysis shows that the \ci(1-0) transition is a reliable alternative to the CO(1-0) line to study the molecular gas properties of the low-excitation gas.

\medskip

\item The CO SLED of MD94 shows excitation properties similar to BX610, another massive galaxy at $z\approx2$ with a comparable SFR surface density ($\Sigma_{\rm SFR}\approx2~M_{\odot}$~yr$^{-1}$~kpc$^{-2}$). Compared to nearby systems, the CO SLED shape is closer to that of Mrk231 ($\Sigma_{\rm SFR}\approx30\times\Sigma_{\rm SFR,MD94}$) than M82 ($\Sigma_{\rm SFR}\approx10\times\Sigma_{\rm SFR,MD94}$), perhaps because MD94 also hosts an AGN. The \ci\ SLED of MD94 is also comparable to that of BX610 and the mean \ci\ SLED of nearby starburst galaxies \citep{israel2015}.

\medskip

\item Based on LVG models from RADEX, we simultaneously fit the CO and \ci\ SLEDs of MD94. The model that best reproduce the data has a gas density of $n_{\rm H_2}=8.5\times10^{3}~{\rm cm}^{-3}$ and a temperature of $T_{\rm kin}=41$~K. These gas properties are similar to those observed in the high-excitation, dense molecular gas in main-sequence galaxies at $z\sim1$ \citep{valentino2020}.

\medskip

\item MD94 exhibits molecular gas properties (e.g., $f_{\rm gas}\approx1$, $t_{\rm dep}\approx0.5$~Gyr) similar to those observed in main-sequence star-forming galaxies at $z\sim2$ \citep[e.g.][]{tacconi18}. The fact that the deconvolved sizes of the disk in the CO(7-6) and CO(1-0) line emission are comparable, suggests that the star formation activity that heat the gas is widely distributed, and not concentrated in a compact central region. Future observations that can spatially resolve the molecular gas properties of MD94 are needed to further confirm this scenario. 

\end{itemize}

\begin{acknowledgements}
We thank the referee for very useful comments and suggestions
that improved the manuscript. K.H.B. and R.H.-C. thanks the Max Planck Society for support under the Partner Group project "The Baryon Cycle in Galaxies" between the Max Planck for Extraterrestrial Physics and the Universidad de Concepción. K.H.B. and R.H.-C. also would like to thank C. Lefèvre, C. Herrera and the rest of the IRAM staff for their help with the data calibration. This work is based on observations carried out under project number W17DL with the IRAM NOEMA Interferometer. IRAM is supported by INSU/CNRS (France), MPG (Germany) and IGN (Spain).
R.H.-C. and R.D. gratefully acknowledges support from the Chilean Centro de Excelencia en Astrof\'isica y Tecnolog\'ias Afines (CATA) BASAL grant AFB-170002. S.B. and R.H-C acknowledge financial support from Millenium Nucleus
NCN19058 (TITANs). M.R. acknowledge support from ANID (Chile) Fondecyt grant No 1190684.
\end{acknowledgements}

%
%



  

\medskip

\bibliographystyle{aa}
\bibliography{referencias_khb.bib}

\end{document}